\documentclass[default,iicol]{sn-jnl}

\begin{document}

\title[Strong Field Ionization of H$_2$]{Strong Field Non-Franck-Condon Ionization of H$_2$: A Semi-Classical Analysis}

\author*[1,2]{
\fnm{Jean-Nicolas}
\sur{Vigneau}}
\email{jean-nicolas.vigneau@universite-paris-saclay.fr}

\author[1]{
\fnm{Osman}
\sur{Atabek}}

\author*[2]{
\fnm{Thanh-Tung}
\sur{Nguyen-Dang}}
\email{thanh-tung.nguyen-dang@chm.ulaval.ca}

\author*[1]{
\fnm{Eric}
\sur{Charron}}
\email{eric.charron@universite-paris-saclay.fr}

\affil[1]{
\orgname{Université Paris-Saclay, CNRS},
\orgdiv{Institut des Sciences Moléculaires d'Orsay},
\orgaddress{
\postcode{91405},
\city{Orsay},
\country{France}}}

\affil[2]{
\orgdiv{Département de chimie, COPL},
\orgname{Université Laval},
\orgaddress{
\street{1045, av. de la Médecine},
\city{Québec},
\postcode{G1V 0A6},
\state{Québec},
\country{Canada}}}

\abstract{Single ionization of H$_2$ molecules exposed to strong and short laser pulses is investigated by a semi-classical method. Three laser characteristics are   considered: i) The carrier-wave frequency corresponds to wavelengths covering and bridging the two ionization regimes: From tunnel  ionization (TI) at 800\,nm to multiphoton ionization (MPI) at 266\,nm. ii) Values of the peak intensity are chosen within a window to eliminate competing double ionization processes. iii) Particular attention is paid to the polarization of the laser field, which can be  linearly  or circularly polarized. The results and their interpretation concern two observables, namely the end-of-pulse total  ionization probability and   vibrational distribution generated in the cation H$_2^+$. The most prominent findings are an increased ionization efficiency in circular polarization and a vibrational distribution of the cation that favors lower-lying levels than those that would be  populated in a    vertical (Franck-Condon) ionization, leading to non Franck-Condon distributions, both in linear and circular polarizations.}


\maketitle

\section{Introduction}
\label{intro}

Imaging and controlling  correlated electronic and nuclear dynamics in their natural time scales is a difficult challenge of ultrafast science \cite{corkum_attosecond_2007}. This is because irradiating a molecule with intense, short laser pulses will lead to a competition between ionization and dissociation processes. Even though these dynamics are expected to evolve on different time scales, a strong field phenomenon such as Coulomb explosion can bring them together and in competition, especially in multi-electron systems. Experimental observations has been reported recently \cite{lu_high-order_2018} of molecular above-threshold dissociation (ATD) during the time scale of the strong-field ionization of dihydrogen, giving rise to concomitant Above-Threshold Ionization spectra (ATI).   Experimental spectroscopic and imaging techniques based  on   the celebrated three-step rescattering mechanism \cite{Corkum-PhysRevLett.71.1994(1993),Lewenstein-PhysRevA.49.2117(1994)}, such as attosecond pump-probe spectroscopy \cite{okino_pump-probe_2015}, high-order harmonic spectroscopy \cite{PhysRevA.82.043414} and laser-induced electron diffraction \cite{Zuo-ChemPhysLett.259.313(1996), Blaga-Nature.483.194(2012), Peters-PhysRevA.83.051403(2011), Mol_Phys_LIED_symmetry_doi:10.1080/00268976.2017.1317858}
can unravel the intricate interplay of electronic and nuclear motions, but they depend on its understanding and control for an optimal functioning of these  as ultrafast analytical tools. 

From a theoretical and computational  viewpoint, a full quantum description of dissociative ionization of multi-electron molecules taking into account electron correlation as well as nuclear motions is highly desirable, and represents a huge technical challenge, scarcely attempted \cite{patchkovskii_high-order-harmonic-spectro_2017} in the strong-field litterature. Even for the simplest many-electron molecular system H$_2$, solving the full multi-dimensional time-dependent Schr\"odinger equation (TDSE) for all electronic and nuclear degrees of freedom represents a formidable computational task. For this  molecule, reduced-dimension models have been used, and allow  partial captures of the coupled electron-nuclear dynamics and even of essential features of the double ionization of the molecule\;\cite{Pegarkov_1999, ADB_HZLu_JPhysB38(2005), PhysRevLett.98.253003, PhysRevA.77.023404}.

No matter the approach one has in mind for the electronic part of the laser-driven molecular  dynamics, it is the description of  ionization steps, strong bound-to-free transitions, that is limiting, and resource-demanding. In view of this, there is an interest to  adapt existing analytical theories of ionization, in particular semi-classical theories such as the Ammosov-Delone-Krainov (ADK) \cite{ammosov_delone_krainov_adk_1986,fabrikant_gallup_adk_2009} and the Perelomov-Popov-Terent'ev (PPT) \cite{perelomov_popov_terentev_ppt_1966, perelomov_popov_terentev2_ppt_1966, perelomov_popov3_ppt_1967} models, into an algorithm for  the numerical calculations of ionization amplitudes out of an arbitrary initial molecular orbital (MO). The ultimate aim would be to incorporate this simplified ionization algorithm into a code treating the full laser-induced electronic dynamics, coupled to vibrational degrees of freedom. There is thus a need to explore how the ionization rate expressions derived in these semi-classical theories can be used to calculate and understand the ionization dynamics of the molecule at a fixed nuclear geometry, (or in a range of nuclear geometries) considered as a parameter, before including  these as parts of a complete quantum dynamical model.

The present theoretical work is  such a preliminary study  devoted specifically to the ionization of H$_2$ molecules exposed to moderately intense laser pulses, investigated by the ADK and PPT  approximations in their adaptations to molecular systems to give the MO-ADK\,\cite{tong_zhao_lin_moadk_2002} and MO-PPT\,\cite{benis-chang-al_moppt_2004} theories respectively. Concentrating on the first ionization of the two-electron molecule, we consider two electronic states: the initial ground state $X^1\Sigma_g^+$ of H$_2$ and the $X^2\Sigma_g^+$ $(1s\sigma_g)$ state of H$_2^+$. More specifically, we are interested in the initial preparation of the cation, in a superposition of its vibrational levels, by an ionizing laser pulse. It is the subsequent dissociation process of the cation that involves the field coupling of the $1s \sigma_g$ and $2p \sigma_u$ states of H$_2^+ $, and this post-ionization dynamics is not considered here. The radiative coupling to the excited $2p\sigma_u$ state of the cation is therefore neglected. Still another simplification we will eventually make is the neglect of the double ionization process, leading to the Coulomb Explosion of the molecule. This is justified if we stay in a rather moderate laser intensity range, the double ionization entering into competition only for intensities exceeding certain thresholds (above a few $10^{14}$\,W/cm$^2$ at 800\,nm, to fix the ideas). The effect of various parameters of the laser field, in particular its  wavelength (in the IR and UV spectral regions), peak intensity and polarization, is analysed in terms of two calculated  observables: The total ionization probability, and the distribution of populations on the vibrational states of the field-free cation, all taken at the final time of the laser pulse. The last observable highlights deviations from the usually assumed Franck-Condon distribution. We will discuss the origin of this non-Franck-Condon distribution in an assessment of the semi-classical ionization model used.  

The manuscript is organized as follows. Section\;\ref{method} starts with the equations governing the time-evolution of the populations of the two-electron molecule in the basis of relevant vibronic states of the neutral H$_2$, the cation H$_2^+$, and the Coulomb Explosion channel H$^{+}$\,+\,H$^{+}$. In physical terms, they are kinetic rate equations describing the relative population gains and losses among these states during the single and double ionization processes. The ionization rates driving these equations have the form of rate constants, and are obtained from the MO-ADK or the MO-PPT theories, and carry a parametric dependence on the internuclear distance $R$ of the molecule through the molecular ionization potential. Section\;\ref{result} gives the results concerning the two observables of interest, discussed in terms of the distinction between tunnel or multiphoton ionization regimes. The most important conclusions of this section are: (i)\;For a given peak intensity, the circularly  polarized field is more efficient than a linearly one in terms of the ionization yield. (ii)\;The molecular ion  vibrational distribution as prepared by the ionization step is shifted from $v_+=2$ (as predicted by a vertical Franck-Condon preparation) to lower vibrational $v_+$ levels. Moreover, when comparing the respective roles of the laser polarization, the differences between linear and circular polarizations in terms of non-Franck-Condon ionization are only noticeable in the tunneling regime.

\begin{figure}[t!]
\centering
\includegraphics[width=0.99\linewidth]{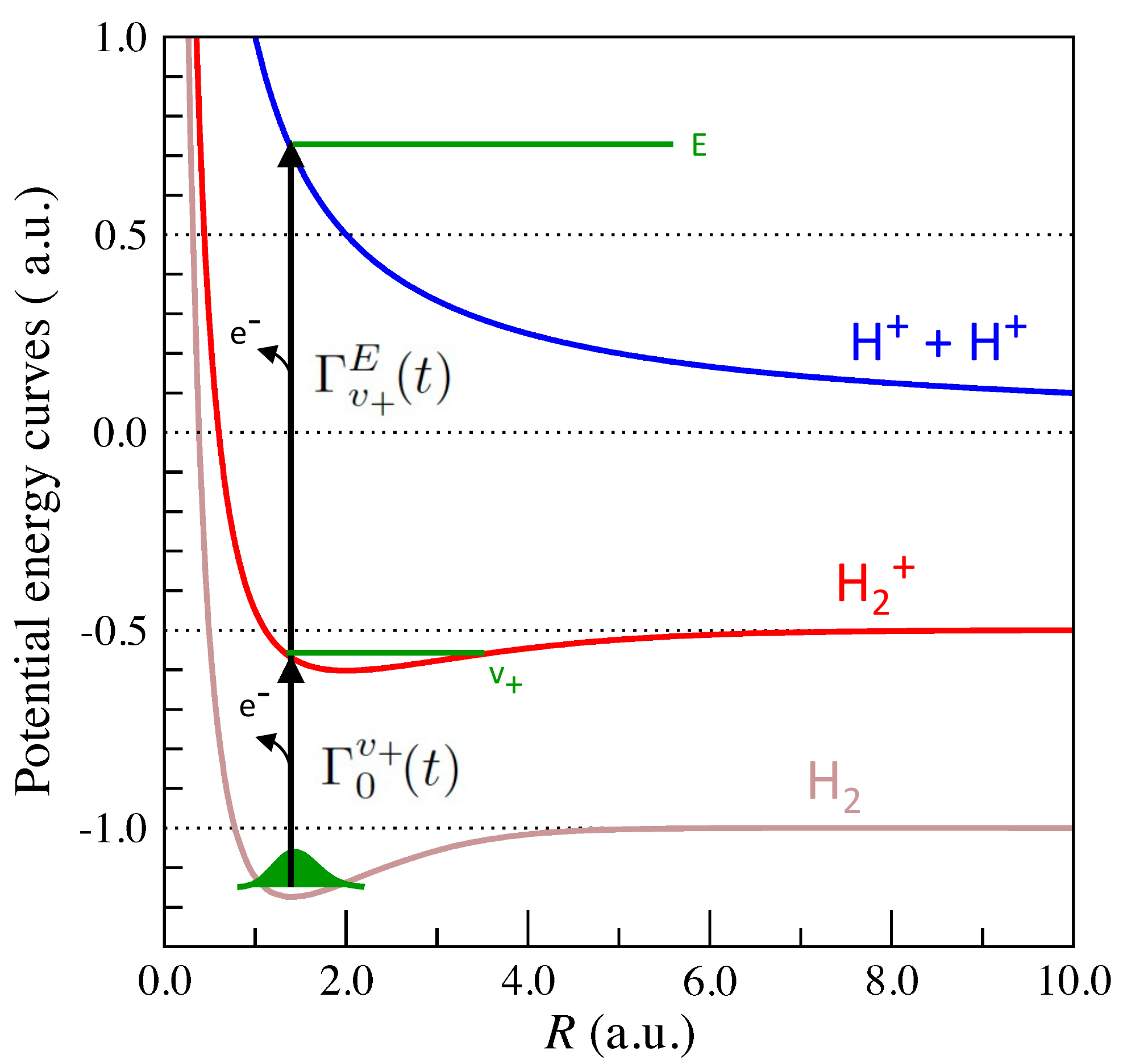}
\caption{{Depiction of the ionization paths of the H$_2$ vibrational state $v=0$ (wave function depicted in green) from its initial potential well (brown) into its ionized vibrational states $v_+$ (green) in the H$_2^+$ ground electronic potential (red), and ultimately to its Coulomb explosion potential H$^+$+H$^+$ (blue) with the kinetic energy $E$ (green).}}
\label{fig:potentials}
\end{figure}

\section{Methods}
\label{method}

The dynamics of single and double ionization of H$_2$ in a time-dependent laser field $F(t)$ is treated by solving the rate equations
\begin{subequations}
\begin{align}
\left.\frac{dP_{\mathrm{H}_2}}{dt}\right\vert_t   & = -\Gamma_0^{+}\!(t)\,P_{\mathrm{H}_2}\!(t)
\label{pop_h2}\\
\left.\frac{dP_{v_+}}{dt}\right\vert_t & = \Gamma_0^{v_+}\!(t)\,P_{\mathrm{H}_2}\!(t)
-\Gamma_{v_+}^{\mathrm{CE}}(t)\,P_{v_+}\!(t)
\label{pop_v+}\\
\left.\frac{dP_{\mathrm{CE}}}{dt}\right\vert_t   & = \sum_{v_+}\,\Gamma_{v_+}^{\mathrm{CE}}(t)\,P_{v_+}\!(t)
\label{pop_c}
\end{align}
\end{subequations}
where $P_{\mathrm{H}_2}\!(t)$ is the total population of H$_2$, $P_{v_+}\!(t)$ is the population of the vibrational level $v_+$ in the ground electronic state of H$_2^+$ and $P_{\mathrm{CE}}(t)$ is the total population in the [H$^+$\,+\,H$^+$] (Coulomb Explosion) dissociation continuum. As a consequence, the total population of H$_2^+$, expressed as
\begin{equation}
P_{\mathrm{H}_2^{+\!}}(t)=\sum_{v_+}\,P_{v_+}\!(t),
\label{pop_h2+}
\end{equation}
is calculated within the state-resolved complete ionization dynamics of the molecule, which includes both the gain of population of the cation from the (first) ionization of the neutral H$_2$ molecule, and its loss due to the second ionization. {The passage of the population from H$_2$ to its ionized states is represented schematically in Fig.\,\ref{fig:potentials}.}

The instantaneous molecular ionization rates $\Gamma_0^{+}\!(t)$, $\Gamma_0^{v_+}\!(t)$ and $\Gamma_{v_+}^{\mathrm{CE}}(t)$ are obtained as described hereafter using the molecular ADK or PPT approximations. The total single ionization rate $\Gamma_0^{+}\!(t)$ is expressed as a sum over all accessible vibrational levels $v_+$ as

\begin{equation}
\Gamma_0^{+}\!(t) = \sum_{v_+}\,\Gamma_0^{v_+}\!(t)
\label{single_ion_rate}
\end{equation}
with
\begin{equation}
\Gamma_0^{v_+}\!(t) = 
\left\vert\int\!\chi_{v_+}^*\!(R)\Big(W_{\mathrm{H}_2}[R,t]\Big)^{\!\!\frac{1}{2}}\!\chi_0(R)\,dR\,\right\vert^2
\label{state_selected_single_ion_rate}
\end{equation}
where $W_{\mathrm{H}_2}[R,t]$ is the instantaneous molecular ADK or PPT ionization rate at the internuclear distance $R$ and at time $t$ given by
\begin{equation}
W_{\mathrm{H}_2}[R,t] = W_{{\mathrm{ADK}}/{\mathrm{PPT}}}\Big[I_{\mathrm{H}_{2\!}}(R),F(t)\Big]\,.
\label{eq:rate_ADK_or_PPT}
\end{equation}
The molecular ADK and PPT ionization rates $W_{\mathrm{ADK}}$ and $W_{\mathrm{PPT}}$ used here are given in Eqs.\,(\ref{moadk}) and\,(\ref{moppt}), and are calculated for the H$_2$ ionization potential $I_{\mathrm{H}_{2\!}}(R)$ and the electric field amplitude $F(t)$. The ionization potential is taken as the energy separating the ground electronic states of H$_2$ and H$_2^+$. In Eq.\,(\ref{state_selected_single_ion_rate}), $\chi_0(R)$ is the wave function associated with the ground vibrational state of H$_2$ and $\chi_{v_+}(R)$ is the wave function associated with the vibrational level $v_+$ of the ground electronic state of the cation. These are calculated numerically using a Numerov type of algorithm\,\cite{Numerov}, to solve for eigenstates supported by the potential energy curves for the ground electronic states of H$_2$ and H$_2^+$, themselves taken from Refs.\,\cite{kolos-wolniewicz_h2-potential_1965} and \cite{peek_h2p-potential_1965, madsen-peek_h2p-potential_1970} respectively.

Similarly, the total ionization rate of the $v_+$ vibrational level of the cation H$_2^+$, $\Gamma_{v_+}^{\mathrm{CE}}(t)$, is expressed as a sum over all Coulomb Explosion energies $E$ as
\begin{equation}
\Gamma_{v_+}^{\mathrm{CE}}(t) = \int\,\Gamma_{v_+}^{E}\!(t)\,dE
\label{double_ion_rate}
\end{equation}
with
\begin{equation}
\Gamma_{v_+}^{E}\!(t) = 
\left\vert\int\!\chi_{E}^*(R)\Big(W_{\mathrm{H}_2^+}[R,t]\Big)^{\!\!\frac{1}{2}}\!\chi_{v_+}\!(R)\,dR\,\right\vert^2
\label{state_selected_double_ion_rate}
\end{equation}
where
\begin{equation}
W_{\mathrm{H}_2^+}[R,t] = W_{{\mathrm{ADK}}/{\mathrm{PPT}}}\Big[I_{\mathrm{H}_2^+}\!(R),F(t)\Big]
\end{equation}
and where $\chi_{E}(R)$ represents the wave function of the energy normalized Coulomb Explosion dissociative continuum at energy $E$. It is calculated numerically using the Numerov algorithm also, with the Coulomb $1/R$ repulsion as potential energy curve. The ionization potential $I_{\mathrm{H}_2^+}\!(R)$ is taken here as the energy separating the $1/R$ Coulomb repulsion energy potential curve from the ground electronic potential curve of H$_2^+$.

At time $t=0$, the system is in the ground vibrational level of the ground electronic state of H$_2$, with
\begin{eqnarray}
P_{\mathrm{H}_{2\!}}(0) & = & 1,\nonumber\\
\forall v_+, \quad P_{v_+}(0) & = & 0, \label{pop_init}\\
P_{\mathrm{CE}}(0)   & = & 0,\nonumber
\end{eqnarray}
and since the sum of Eqs. (\ref{pop_h2}), (\ref{pop_v+}) and (\ref{pop_c}) is zero, one can note that the total population 
\begin{equation}
P_{\mathrm{H}_{2\!}}(t)+P_{\mathrm{H}_2^{+\!}}(t)+P_{\mathrm{CE}}(t)=1
\label{pop_tot}
\end{equation}
is naturally conserved throughout the entire ionization process.

\begin{figure}[t!]
\centering
\includegraphics[width=0.99\linewidth]{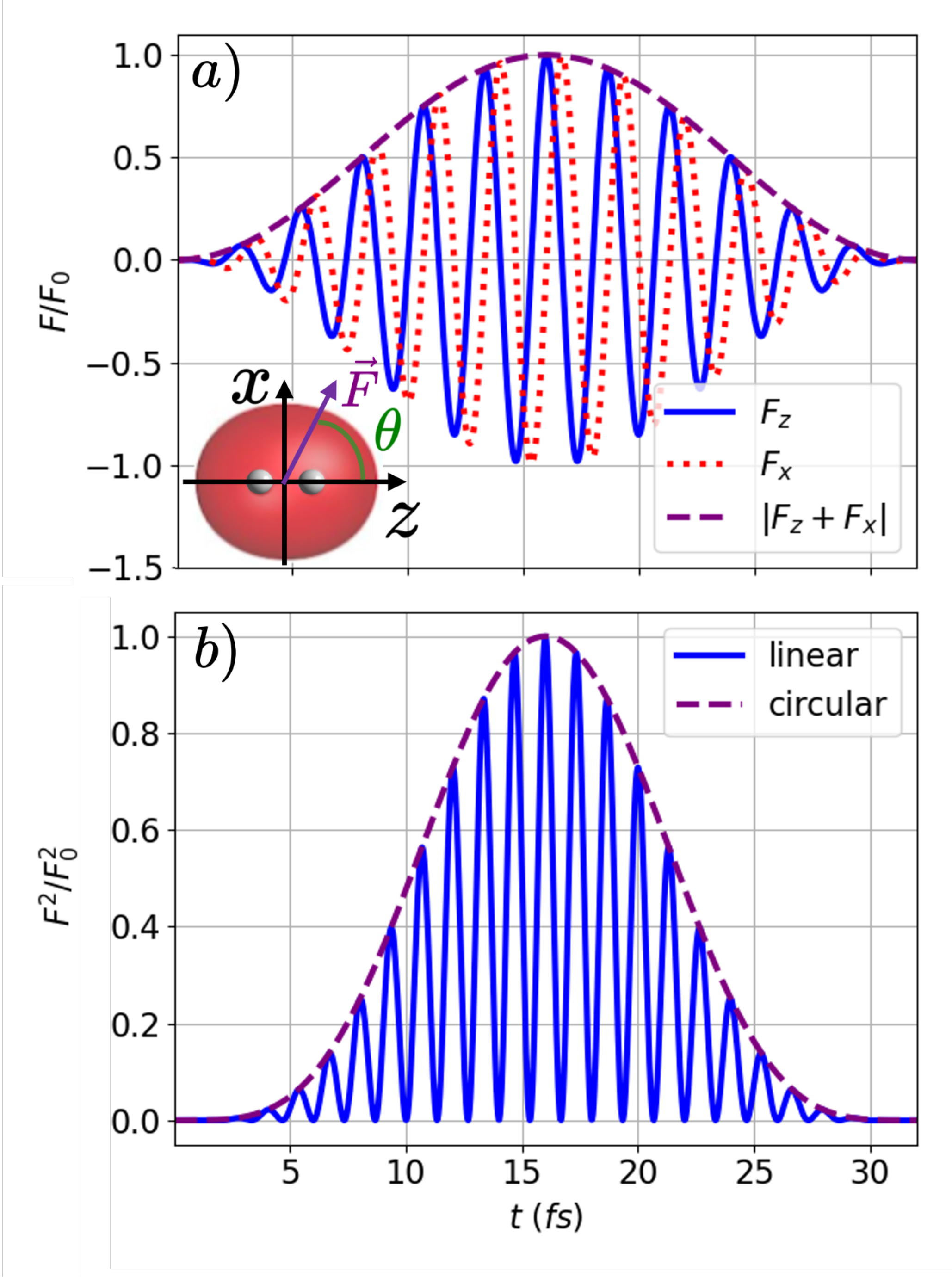}
\caption{Panel (a): 800\,nm circularly polarized laser electric field components along the $z$-axes ($F_z(t)$, full blue line), and along the $x$-axes ($F_x(t)$, dotted red line) as a function of time. The H$_2$ molecular axis is aligned along $z$-axis. $\theta$ is the angle between the intermolecular axis and the instantaneous field direction. Panel (b): Square modulus of the linear (full blue line) and circular  (dashed purple line) fields as a function of time.}
\label{fig:field}
\end{figure}

The molecular ADK (or MO-ADK) ionization rate is given by \cite{tong_zhao_lin_moadk_2002}
\begin{equation}
W_{\mathrm{ADK}}\big[I_p,F\big] = B_0^2 \left(\frac{2\kappa^2}{F}\right)^{\!\!\xi} \exp\!\left(-\frac{2\kappa^3}{3F}\right),
\label{moadk}
\end{equation}
where $\kappa=\sqrt{2I_p}$, $\xi=(2Z/\kappa)-1$, and
\begin{equation}
B_0(\theta) = \sum_l\,\sqrt{\frac{2l+1}{2}}\,C_{l}\,P_l(\cos\theta)\,.
\label{B0}
\end{equation}
$Z$ is the effective Coulomb charge seen asymptotically by the electron and $\theta$ is the angle between the laser field and the molecular axis. In the case of linear polarization, the time dependent electric field amplitude is given by $F(t)=\vert\mathbf{F}(t)\vert$, where
\begin{equation}
\mathbf{F}(t) = F_0\,f(t)\cos(\omega t)\,\mathbf{e}_\theta
\label{linear_field}
\end{equation}
$\omega$ is the carrier-wave frequency and $F_0$ the electric-field amplitude. $f(t)=\sin^2(\pi t/t_f)$ denotes the temporal envelope of the pulse, $t_f$ being the total pulse duration, $\mathbf{e}_\theta=[\cos\theta\,\mathbf{e}_z+\sin\theta\,\mathbf{e}_x]$ is the unit vector along the direction of  the fixed  polarization    of the field, which makes an angle $\theta$ with respect to the molecular  axis. In this case of linear polarization, we will limit ourselves to the angle $\theta=0$, corresponding to a situation where the molecule is aligned with the field. In the case of circular polarization, the time dependent electric field is given by
\begin{equation}
\mathbf{F}(t) = F_0\,f(t)\Big[\cos(\omega t)\,\mathbf{e}_z+\sin(\omega t)\,\mathbf{e}_x\Big]\,,
\label{circular_field}
\end{equation}
i.e. the angle $\theta$ becomes time-dependent, with $\theta(t)=\omega t$. It is to be noted that the field amplitude $F_0$ for the circular polarization case is chosen as the one for the linear case divided by $\sqrt{2}$, such that the average value of $F^2(t)$ over an optical cycle is identical in linear and circular polarization. The coefficients $C_{l}$ in Eq.\,(\ref{B0}) ensure the anisotropy of the ionization by the different weighting of the spherical harmonics-related Legendre polynomials $P_l(\cos\theta)$ of quantum number $l$. The values of the coefficients $C_{l}$ are experimentally fitted in Refs.\,\cite{Awasthi_SAE_2008} and\,\cite{zhao-jin-al_coeff-h2p_2010}, and are provided in Table \ref{table_c-aniso} for both the ionization of H$_2$ and of H$_2^+$. Figure\,\ref{fig:field} shows the electric field components $F_z$ and $F_x$ and their square modulus for circular polarization along the $z$ and $x$ axes. The linear polarization corresponds solely to $F_z$. The upper panel also illustrates the angular positioning of the H$_2$ molecule assumed to lie on the $z$-axis.

\begin{table}[t]
\begin{center}
\begin{minipage}{174pt}
\caption{Anisotropy coefficients $C_{l}$ for H$_2$ \cite{Awasthi_SAE_2008} and H$_2^+$ \cite{zhao-jin-al_coeff-h2p_2010}}\label{table_c-aniso}
\begin{tabular}{@{}ccc@{}}
\toprule
$l$ & $C_{l}$ for H$_2$ & $C_{l}$ for H$_2^+$\\
\midrule
$0$ & $2.4350$ & 4.52 \\
$2$ & $0.1073$ & 0.62 \\
$4$ & $0.0010$ & 0.03 \\
\botrule
\end{tabular}
\end{minipage}
\end{center}
\end{table}

In the molecular PPT (or MO-PPT) approach, the ionization rate is given by \cite{benis-chang-al_moppt_2004}
\begin{align}
W_{\mathrm{PPT}}\big[I_p,F\big] = & B_0^2 \left(\!\frac{2\kappa^2}{F\sqrt{1+\gamma^2}}\!\right)^{\!\!\xi}\! \exp\!\left(\!-\frac{2\kappa^3}{3F}g(\gamma)\!\right)\nonumber\\
    & \times A_0(\omega,\gamma)\left(1+2\gamma\right)^{-2Z/\kappa},
\label{moppt}
\end{align}
when taking into account the Coulomb correction factor \cite{popruzhenko-bauer-al_coul-corr_2008}. In this expression $\gamma=\omega\sqrt{2I_p}/F$ is the (frequency dependent) Keldysh parameter \cite{Keldysh-JETP.20.1307(1965), popruzhenko_keldysh_theory_2014}. The expressions of $g(\gamma)$ and $A_0(\omega,\gamma)$ differ in linear and circular polarizations. For the case of linear polarization, $A_0(\omega,\gamma)$ is given by
\begin{equation}
A_0(\omega,\gamma)=\frac{\beta^2(\gamma)}{\sqrt{3\pi}}\!\sum_{\tilde{n}\gt\nu}^{\infty} \Omega_{0\!}\Big(\!\sqrt{\tilde{n}'\beta(\gamma)}\Big)e^{-\tilde{n}'\alpha(\gamma)}
\label{lin_a_fct}
\end{equation}
where $\tilde{n}'=\tilde{n}-\nu$ with $\nu=\frac{I_p}{\omega}\left(1+\frac{1}{2\gamma^2}\right)$. The sum therefore runs over the number of photons $\tilde{n}$ absorbed above the ionization threshold. In practice, in the numerical evaluation of $A_0(\omega,\gamma)$, we considered the sum (\ref{lin_a_fct}) converged when its relative gain over an iteration was less than $10^{-10}$ of the current sum value. In addition, we have
\begin{subequations}
\begin{eqnarray}
\label{omega_fct}
\Omega_0(x)    & = & \frac{x}{2} \int_0^1 \frac{e^{-x^2t}}{\sqrt{1-t}} dt\,,\\[0.2cm]
\label{alpha_fct}
\alpha(\gamma) & = & 2\operatorname{Arsh}(\gamma)-\beta(\gamma)\,,\\[0.2cm]
\label{beta_fct}
\beta(\gamma)  & = & \frac{2\gamma}{\sqrt{1+\gamma^2}}\,,\\
g(\gamma)      & = & \frac{3}{2\gamma}\Big[\!\Big(\!1\!+\!\frac{1}{2\gamma^2}\!\Big)\!\operatorname{Arsh}(\gamma)\!-\! \frac{1}{\beta(\gamma)}\Big]
\end{eqnarray}
\end{subequations}

Finally, in the case of circular polarization $A_0$ is a a function of $\gamma$ only, and the expressions of $A_0(\gamma)$ and $g(\gamma)$ involved in the MO-PPT ionization rate (\ref{moppt}) become
\begin{equation}
A_0(\gamma)=(1\!-\!t_0)\!\left[\frac{\gamma^2(1\!+\!\gamma^2)(1\!-\!t_0^2)}{(\gamma^2\!+\!t_0^2)(1\!+\!t_0^2\!+\!2t_0^2/\gamma^2)}\right]^{\frac{1}{2}}
\label{circ_a_fct}
\end{equation}
and
\begin{equation}
g(\gamma)=\frac{3t_0}{\gamma^2(1-t_0^2)}\left[(1+\gamma^2)\left(1+\frac{t_0^2}{\gamma^2}\right)\right]^{\frac{1}{2}}
\label{circ_g_fct}
\end{equation}
where $t_0$ function of $\gamma$, is a real positive root, in the interval [0,1], of the following equation \cite{perelomov_popov_terentev_ppt_1966}
\begin{equation}
\label{t0}
\operatorname{Arth}\left(\sqrt{\frac{t_0^2+\gamma^2}{1+\gamma^2}}\right) = \frac{1}{1-t_0}\sqrt{\frac{t_0^2+\gamma^2}{1+\gamma^2}}\,.
\end{equation}
For practical reasons, it is convenient to fit $t_0(\gamma)$ using a simple analytical function. Indeed, one can very accurately fit this variation with $\gamma$ using an eighth order Padé approximant \cite{Pade1892} in the form
\begin{equation}
t_0(\gamma) = \left( \sum_{i=0}^{8} a_i\,\gamma^i \right) \bigg/ \left( \sum_{i=0}^{8} b_i\,\gamma^i \right)
\label{t0_pade}
\end{equation}
The corresponding fitting parameters $a_i$ and $b_i$ are provided in table \ref{tab_pade}. We have verified that the relative error made when using this simplified analytic expression is always less than $6 \times 10^{-4}$ on the interval $\gamma \in [0,100]$.

\begin{table}[ht]
\begin{center}
\begin{minipage}{174pt}
\caption{Padé approximant parameters for $t_0$.}
\label{tab_pade}
\begin{tabular}{@{}ccc@{}}
\toprule
$i$ & $a_i$ & $b_i$ \\
\midrule
0 & 0 & 1 \\
1 & 0 & 1.97454 \\
2 & 1/3 & 2.50994 \\
3 & $6.58161\times10^{-1}$ & 2.46203 \\
4 & $6.29483\times10^{-1}$ & 1.55840 \\
5 & $4.09872\times10^{-1}$ & $7.56738\times10^{-1}$ \\
6 & $1.41151\times10^{-1}$ & $1.97492\times10^{-1}$ \\
7 & $9.97954\times10^{-3}$ & $1.23464\times10^{-2}$ \\
8 & $1.02588\times10^{-4}$ & $1.17511\times10^{-4}$ \\
\botrule
\end{tabular}
\end{minipage}
\end{center}
\end{table}

\section{Results}\label{result}

In this section we first justify the consideration of the first step of the complete ionization dynamics of H$_2$, the first ionization of the two-electron molecule,  by referring to an intensity range which allows one to separate its single and double ionizations. We then analyze the performances of the two semi-classical approaches, MO-ADK and MO-PPT, by comparing the ionization yields they predicted to results of  full quantum mechanical TDSE calculations. We confirm through this comparison that MO-PPT is actually a rather good approximation for describing ionizations both in tunnel or multiphoton regimes. Finally, we discuss and interpret results of MO-PPT calculations on   two observables, namely the final ionization probability vs. intensity profiles and the non Franck-Condon vibrational distributions of H$_2^+$ as a function of the field intensity and polarization characteristics.

\subsection{Single versus double ionization}\label{3.1}

The laser-induced complete dissociative ionization of H$_2$, in a full quantum description, should consider the two electrons and the two protons on an equal footing, through a wavefunction evolution taking electron correlation fully into account and  involving both the single and double ionization processes, together with the nuclear dynamics. Such a full quantum treatment remains a formidable methodological challenge. Depending on the laser pulse duration and leading intensity, the complete dissociative ionization dynamics can be described by steps of increasing sophistication. In particular, the two time scales of ionization and nuclear  dynamics being very different, the possibility is offered to treat separately the two fragmentation processes, focusing as a first step solely on the ultra-fast ionization. Depending on the intensity of the laser, this will give rise to single, and sequential (or non-sequential) double ionization processes \cite{Walker-PhysRevLett.73.1227(1994), PhysRevA.77.023404, Mauger-PhysRevLett.108.063001(2012)}. Neglecting all vibrational wave packet evolution during the pulse, all the ultra-fast ionization dynamics is encapsulated in the rate $W_{\mathrm{H}_2}[R,t]$ appearing in Eq.\,(\ref{state_selected_single_ion_rate}) which, as stated above, is obtained  within a semi-classical approach  either the simple, commonly used MO-ADK approach \cite{tong_zhao_lin_moadk_2002} or the more complete MO-PPT theory \cite{benis-chang-al_moppt_2004}.

\begin{figure}[ht!]
\centering
\includegraphics[width=0.99\linewidth]{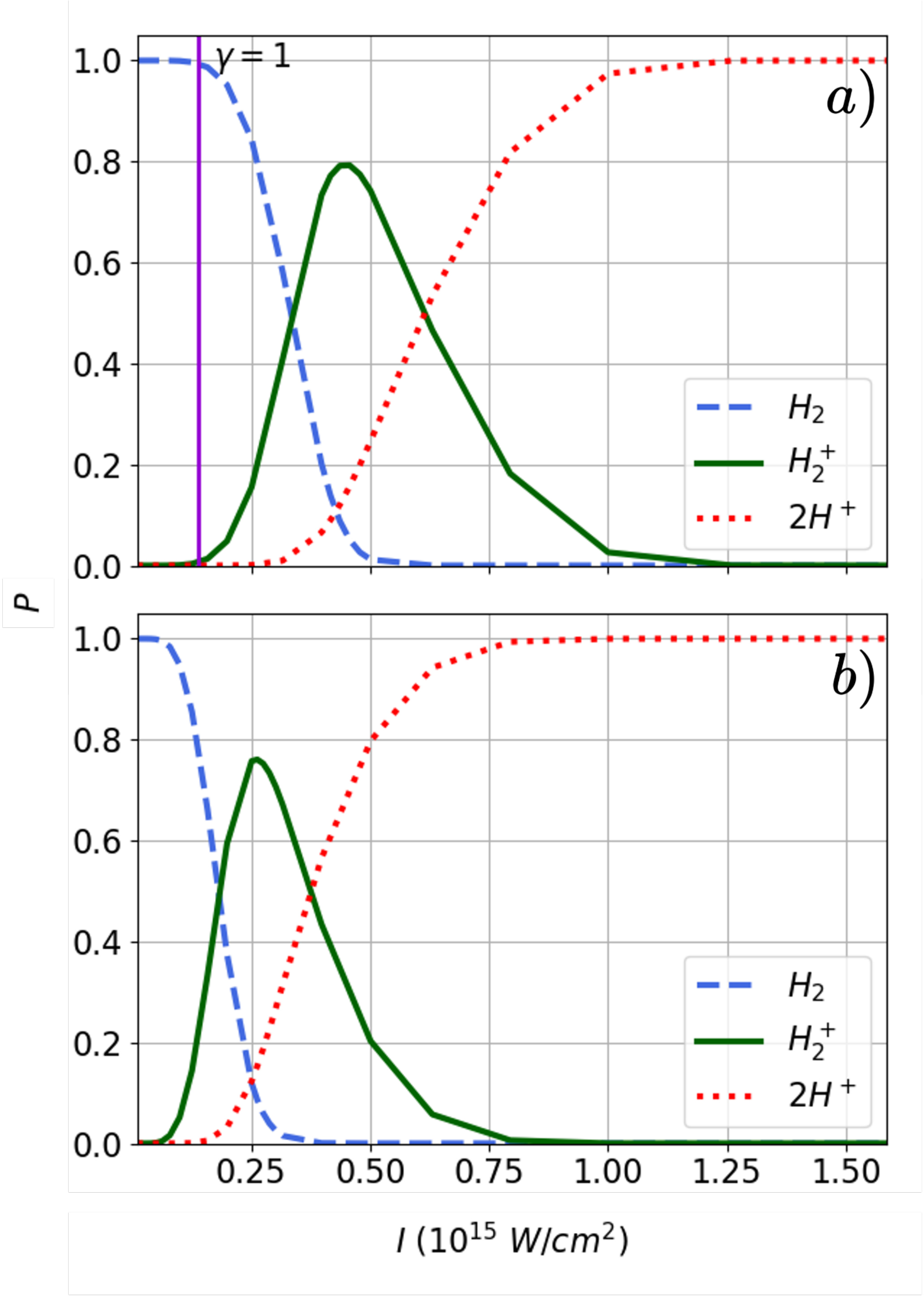}
\caption{Final populations of H$_2$ ($P_{\mathrm{H}_{2\!}}(t_f)$; dashed blue line), H$_2^+$ ($P_{\mathrm{H}_2^{+\!}}(t_f)$; full green line) and H$^+$\,+\,\,H$^+$ ($P_{\mathrm{CE}}(t_f)$; Coulomb Explosion, red dotted line), using a 800\,nm $(a)$ linearly and $(b)$ circularly polarized field, and the MO-PPT approach {for a sin$^2$ pulse envelope of total duration 32\,fs corresponding to 12 optical cycles (16\,fs FWHM)}. The vertical purple line in panel (a) indicates the value of the intensity at which the Keldysh parameter is $\gamma=1$.}
\label{fig:pop}
\end{figure}

Figure \ref{fig:pop} illustrates how the final  population, (evaluated at the final time $t_f$ of the laser pulse), of the initial neutral H$_2$ molecule, $P_{\mathrm{H}_2}\!(t_f)$, that of the molecular cation H$_2^+$ resulting from the first ionization, $P_{\mathrm{H}_2^+}\!(t_f)$, and the population of the Coulomb Explosion channel accessed from the second ionization, $P_{\mathrm{CE}}(t_f)$, vary as a function of the laser peak intensity $I$. For this illustrative purpose we are showing the results for a $\lambda=800$\,nm linearly or circularly polarized  IR pulse extending over 12 optical cycles, for a total duration of about 32 fs. {Results obtained by doubling or halving the pulse duration show that changing the pulse duration  only scales up or down the total ionization probability without changing the relative populations of the different vibrational levels produced in H$_2^+$. These  results are collected in the Supplementary Information.}

In Fig.\,\ref{fig:pop}, the first ionization is observed for peak intensities of about $10^{14}$\,W/cm$^2$, with the precise onset intensity depending on the polarization of the field. There, the initial population of the neutral molecule starts  to decrease while the H$_2^+$ cation's  population increases up to about 0.8. It is only for field intensities typically larger than $4 \times 10^{14}$\,W/cm$^2$ in linear polarization and $2.5 \times 10^{14}$\,W/cm$^2$ in circular polarization that the second ionization process comes into play and the initial state population decreases below 0.2. {Note that this onset of second ionization would be shifted to a higher intensity for a shorter pulse (see Supplementary Information for details).} It is worthwhile to note that this second ionization sets in basically by depleting the population of the H$_2^+$ cation. It is precisely in  that respect that we can at least partly separate the two (first and second) ionization processes, defining the physical frame for our focus on the single ionization as a dominant process for intensities typically less than about $3 \times 10^{14}$\,W/cm$^2$ at 800\,nm. This thus fixes the general framework of the present study which will concern from now on only the first ionization of H$_2$. 

To be complete, a few additional observations are in order: (i) For intensities exceeding $10^{15}$\,W/cm$^2$ the molecule is completely ionized both for linear or circular polarizations. (ii) An identification as to whether the ionization follows a tunnel or multiphoton regime can be given in terms of the already defined Keldysh parameter $\gamma$. The regime with $\gamma \ll 1$ is referred to as the tunnel ionization (TI) regime, where the electron escapes by tunneling through the barrier resulting from a field-distorted Coulomb potential. In the opposite limit of $\gamma \gg 1$, the electron ionizes after absorption of several photons, a regime termed multiphoton ionization (MPI). The field intensity value giving $\gamma=1$ for the first ionization process is indicated by a vertical line in Fig.\,\ref{fig:pop}(a). It roughly delimits the two ionization regimes (TI and MPI). For a given ionization potential and field frequency, the Keldysh parameter takes on values less than 1 for field intensities higher than the limit indicated by the corresponding vertical line. In other words, basically all intensities within the first ionization profile studied here allow at 800\,nm an interpretation of the dynamics in terms of the TI regime, where semi-classical approaches are normally rather well adapted. (iii) Finally, linear and circular polarizations give slightly different results. More precisely, switching the polarization from linear to circular, the first ionization $P_{\mathrm{H}_2^{+\!}}$ vs. $I$ profile is shifted towards lower intensities and acquires a narrower extension.  This could be interpreted by the fact that for the circularly polarized pulse, as opposed to the linearly polarized one, the electric field amplitude does not vanish at any time during the optical cycle. Thus, for a given nominal intensity the circularly polarized field has a higher fluence, and hence a better ionization efficiency.

\subsection{Validation of the semi-classical approaches}

\begin{figure}[t!]
\centering
\includegraphics[width=0.99\linewidth]{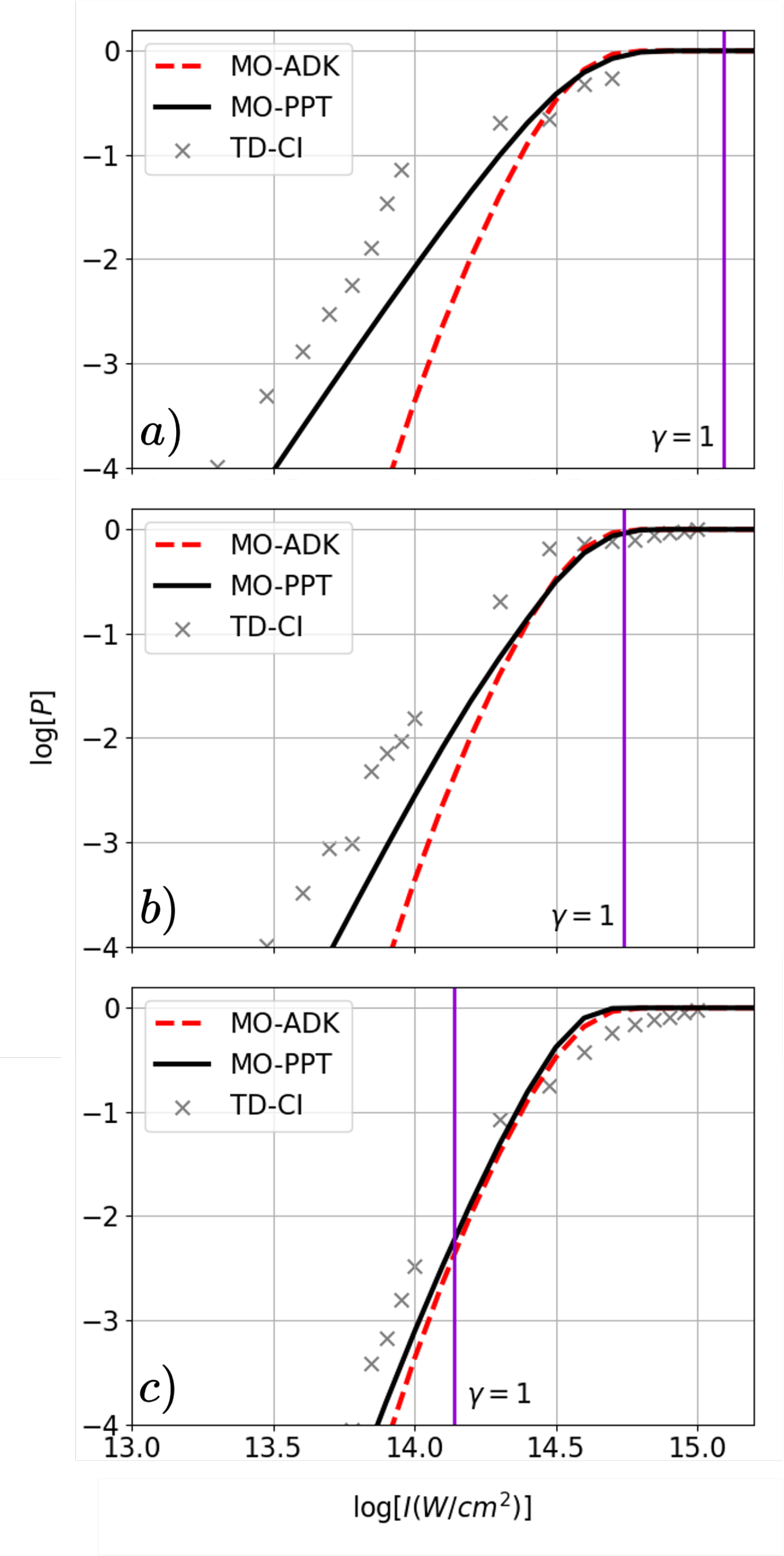}
\caption{Final ionization probability of H$_2$ (in logarithmic scale) using a linearly polarized field of peak intensity $I$ (in logarithmic scale), as given by MO-ADK (dashed red line), MO-PPT (full black line) and TDCI \cite{Awasthi_SAE_2008} (grey crosses) methods, with wavelengths of $(a)$ 266 nm over 36 optical cycles, $(b)$ 400 nm over 24 optical cycles and $(c)$ 800 nm over 12 optical cycles. {The total pulse duration is therefore fixed at the value 32\,fs (16\,fs FWHM).}}
\label{fig:1stion_lin}
\end{figure}

\begin{figure}[t!]
\centering
\includegraphics[width=0.99\linewidth]{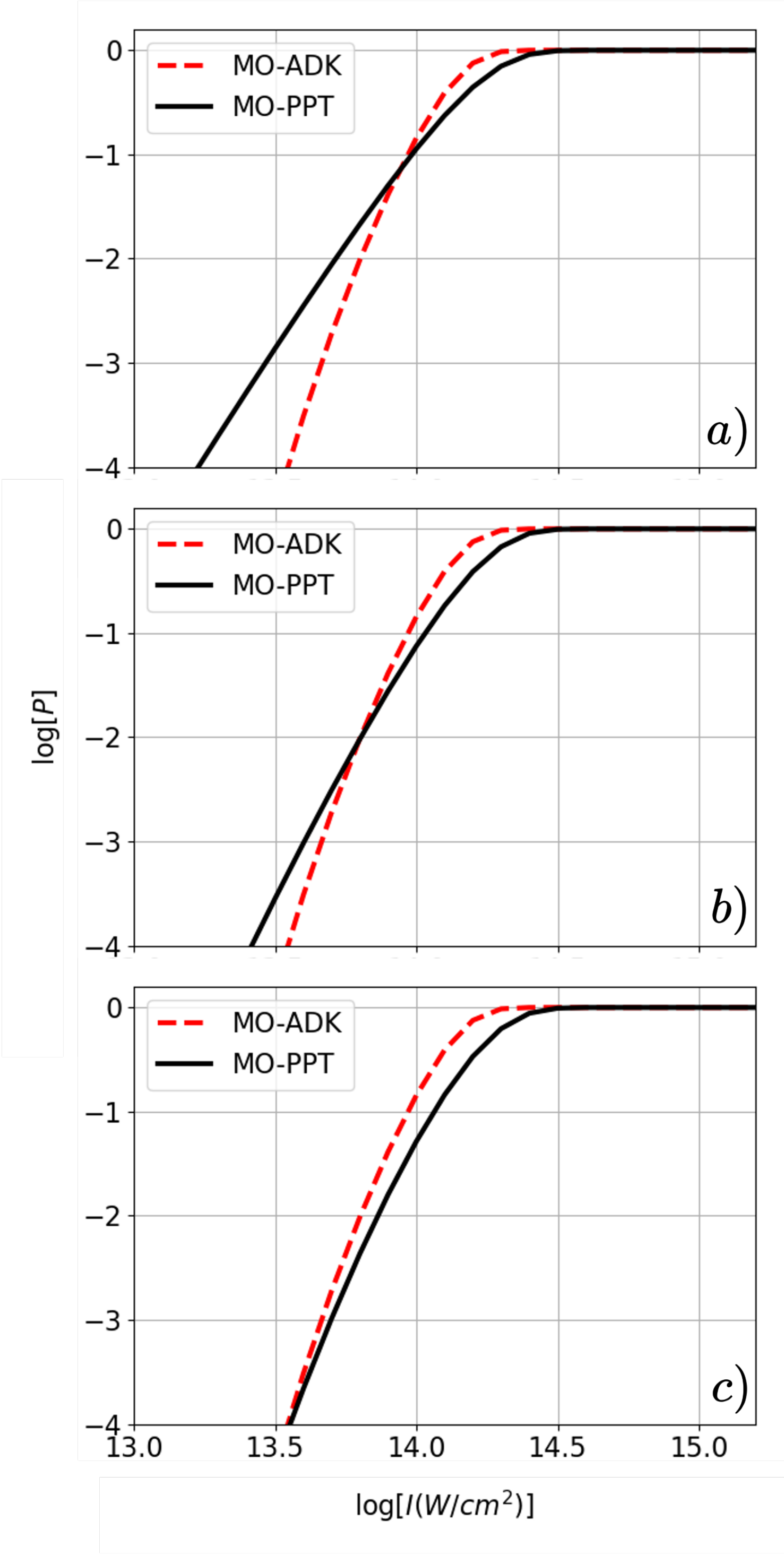}
\caption{Final ionization probability of H$_2$ (in logarithmic scale) using a circularly polarized field of peak intensity $I$ (in logarithmic scale), as given by MO-ADK (dashed red line) and MO-PPT (full black line) methods, with wavelengths of $(a)$ 266 nm over 36 optical cycles, $(b)$ 400 nm over 24 optical cycles and $(c)$ 800 nm over 12 optical cycles. {The total pulse duration is therefore fixed at the value 32\,fs (16\,fs FWHM).}}
\label{fig:1stion_circ}
\end{figure}

We proceed now to a quantitative assessment of the performances of the two semi-classical approximations MO-ADK and MO-PPT with respect to the first ionization of H$_2$, by comparing the calculated $P_{\mathrm{H}_2^{+\!}}(t_f)$ with that predicted by a full quantum mechanical calculation \cite{Awasthi_2005,Vanne_2004,Awasthi_SAE_2008}, in which the two-electron TDSE was solved at a {(full) Time-Dependent
Configuration Interaction (TDCI) level, using CI wavefunctions comprising of thousands of configurations using  Kohn-Sham (DFT) molecular orbitals expressed in an extended B-spline basis. The TDCI data used here are those of figures 3, 5, 7 of Ref.\,\cite{Awasthi_SAE_2008} explicitly marked TDCI (as opposed to SAE-CI, based on a more limited CI expansion of the two-electron system eigenstates, as per the description of it in Ref.\,\cite{Awasthi_SAE_2008}). Note that this paper (Ref.\,\cite{Awasthi_SAE_2008}) assesses the validity of the SAE by comparing SAE calculations' results with the aforementioned TDCI results. It is the latter however that are relevant here}. The comparison is made  for a series of laser peak intensities ranging from $10^{13}$\,W/cm$^2$ up to $10^{15}$\,W/cm$^2$, and for three wavelengths, $\lambda=266$\,nm, 400\,nm, 800\,nm. The number of optical cycles are adjusted, for each value of $\lambda$, in such a manner as to have the same pulse duration. Both  polarization cases, linear and circular, are considered. Figure\,\ref{fig:1stion_lin} displays, for the linear polarization case, the final ionization probability as a function of intensity on a logarithmic scale to better enhance the large scale differences. The corresponding results for circular polarization are displayed in Fig.\,\ref{fig:1stion_circ}.

On numerical grounds, the present semi-classical calculations are in close agreement (graphical estimation) with those of the Fig.\;5 of Ref.\,\cite{PhysRevA.93.023413}, despite a difference in the models for the calculation of the rate: Averaged over H$_2^+$ vibrational states in this work (Eq.\,\ref{state_selected_single_ion_rate}), and taken at H$_2^+$ equilibrium distance for Ref.\,\cite{PhysRevA.93.023413}. We note that for the highest frequency considered ($\lambda=266$\,nm, panel (a) of Fig.\,\ref{fig:1stion_lin}) the ionization is basically in the multiphoton (MPI) regime, as all considered values of the field intensity $I$ up to $10^{15}$\,W/cm$^2$ give a Keldysh parameter $\gamma > 1$. In this case, the frequency-independent MO-ADK description fails by orders of magnitude in producing the correct H$_2^+$ population, at least up to $I=3\times 10^{14}$\,W/cm$^2$. The MO-PPT model is a much better approximation, and gives results that basically follow the exact TDCI results. Semi-classical approximations are in better agreement among them and with the exact results for lower frequencies ($\lambda=400$\,nm, 800\,nm, Fig.\;\ref{fig:1stion_lin}, panels (b), (c) respectively), where the tunnel mechanism becomes progressively dominant. This is indicated by the shift of the vertical line $\gamma=1$ towards lower intensities in panels (b) and (c). In particular, a rather good agreement between MO-ADK and MO-PPT results is obtained at $\lambda=800$\,nm. 

Although no TDCI calculations results are available for the case of circular polarization, we can still offer a reasonable interpretation of two observations in Fig.\;\ref{fig:1stion_circ}: (i) for the three wavelength considered, the threshold intensity for the H$_2^+$ population saturation is lower for the case of a circular polarization as compared to the linear polarization case, due to the higher fluence argument advanced previously; (ii) the slope  of the rise of the ionization probability curve is much higher for $\lambda=800$\,nm than for $\lambda=266$\,nm. This is due to the fact that $\lambda=800$\,nm corresponds to the TI regime where the ionization yield is exponentially increasing with the intensity, as opposed to the $\lambda=266$\,nm case where the MPI regime is the leading one, giving rise to a slower power-type increase.

\subsection{Non-Franck-Condon ionization}

An interesting way of presenting the results of the integration of Eq.\,(\ref{pop_v+}) without the last term on the right hand side (since we are considering only the first ionization process, ignoring the Coulomb Explosion channel) is to plot, as a normalized function of $v_+$, in the form of a histogram, the populations $f^{nFC}(v_+)$ of various vibrational states $v_+$ of the cation, relative to the final total population of cation $P_{\mathrm{H}_2^+}\!(t_f)$, accumulated during the ionizing action of the laser pulse
\begin{eqnarray}
\label{Pv_plus_nonFC}
f^{nFC}(v_+) & = & \frac{P_{v_+}(t_f)}{P_{\mathrm{H}_2^+}\!(t_f)}\nonumber\\
& = & \frac{1}{P_{\mathrm{H}_2^+}\!(t_f)}\int_0^{t_f} \!\!\Gamma_0^{v_+}\!(t)  P_{\mathrm{H}_2}\!(t)dt
\end{eqnarray}
This kind of plot gives the composition of the vibrational wave packet created in the cation as it is formed by the ionization of the neutral molecule H$_2$. It is often assumed that the ionization is a Franck-Condon (FC) process, \emph{i.e.} the vertical promotion of the ground vibrational state $\chi_0(R)$ of H$_2$ onto the cation electronic ground-state manifold. This Franck-Condon principle corresponds to assume that $W_{\mathrm{H}_2}[R,t]$ in Eq.\,(\ref{state_selected_single_ion_rate}) does not depend on the internuclear distance $R$. Eq.\,(\ref{Pv_plus_nonFC}) then becomes
\begin{equation}
\label{Pv_plus_FC}
f^{FC}(v_+) = \frac{\vert\langle\chi_{v_+}\vert\chi_0\rangle\vert^2}{\sum_{v_+}\vert\langle\chi_{v_+}\vert\chi_0\rangle\vert^2}
\end{equation}

Now, by the mere fact that  the rate $W_{\mathrm{H}_2}[R,t]$ as described by the MO-ADK formula for tunnel ionization, or by the more general MO-PPT expression, depends on the internuclear distance $R$ through the first ionization potential $I_{\mathrm{H}_{2\!}}(R)$ (see Eq.\,\ref{eq:rate_ADK_or_PPT}), we expect that the vibrational distribution in the cation does not follow the Franck-Condon principle. Indeed, the onset of a non-Franck-Condon behavior has been related to a sharp increase of the tunnel ionization rate as a function of the internuclear distance $R$ by a number of authors \cite{Saenz_2000, posthumus2001super}. {Such an increase in the ionization rate is expected when $I_p$ decreases, and this is the case when $R$ increases in the range $0 < R < 3.5$\,a.u which is of interest to us, hence the relationship between $\Gamma_{0}^{v_+}$ and $R$.} Another form of deviation from the Franck-Condon principle has been discussed in the literature \cite{PhysRevLett.92.163004}, and pertains more to the field-induced deformation of the cation vibrational states onto which the initial (also deformed) vibrational wavefunction of the parent neutral molecule is projected. The ADK or PPT ionization rates used in this work contain the $R$-dependent ionization potential, and this does contribute in part to the non-Franck-Condon effect demonstrated therein. We also extend the analysis of non-Franck-Condon ionization to circular polarization, and to several wavelengths covering the TI and MPI regimes.

\begin{figure}[ht!]
\centering
\includegraphics[width=0.99\linewidth]{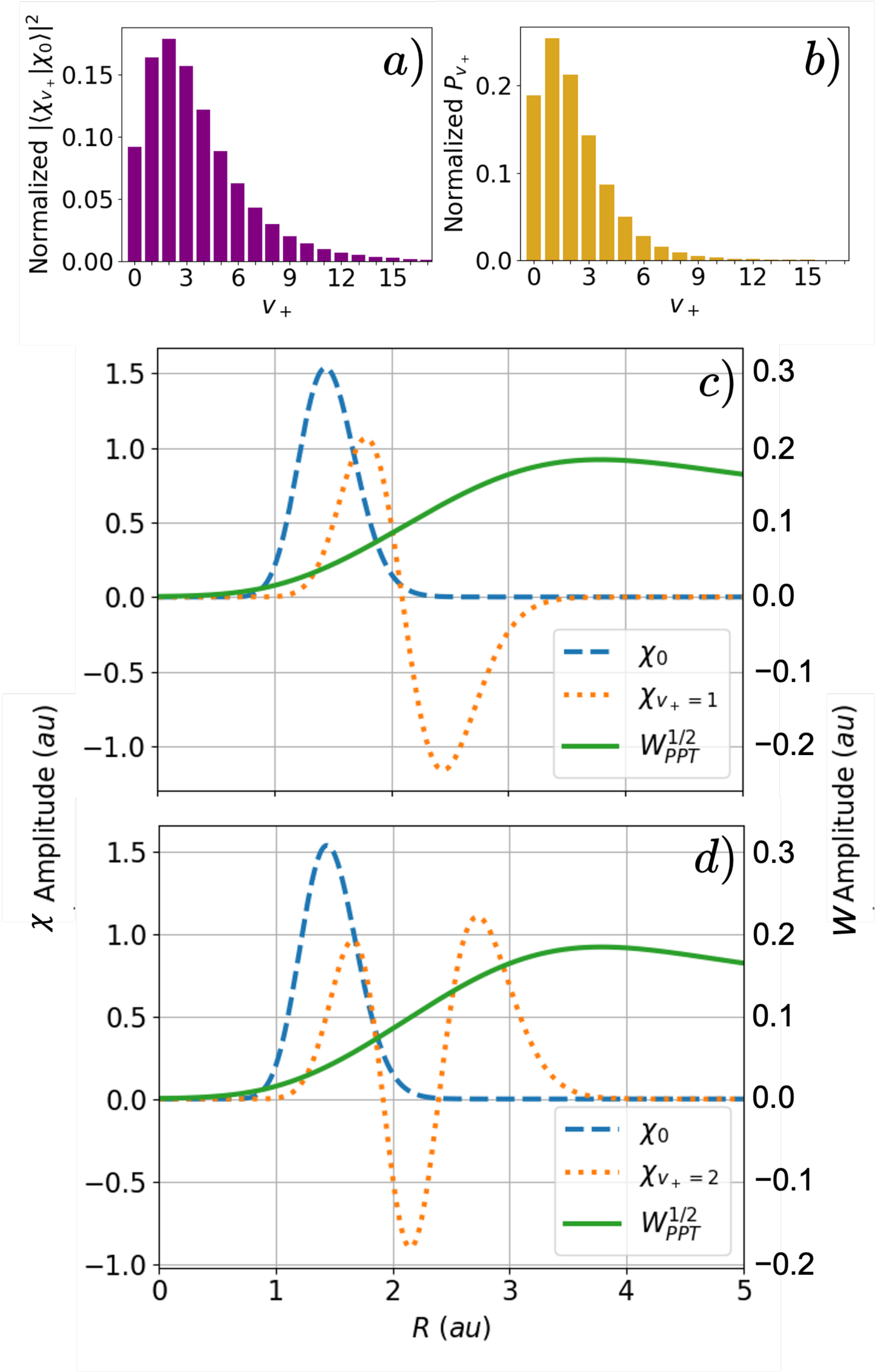}
\caption{Franck-Condon projection of the H$_2$ ground state on the H$_2^+$ vibrational states $v_+$ (upper panel a), and the non-Franck-Condon population resulting from a MO-PPT circularly-polarized 800 nm field of peak intensity $1.6\times10^{14}$\, W/cm$^2$ (upper panel b). The lower panels display the wavefunctions of H$_2$ ground state ($\chi_0$), together with the ones of the vibrational states $(c)$ $v_+=1$ and $(d)$ $v_+=2$ of H$_2^+$ ($\chi_{v_+}$). The square root of the ionization amplitude calculated from the MO-PPT method ($W_{\mathrm{PPT}}^{1/2}$, from Eq.\,\ref{moppt}) using a 800 nm circularly polarized field and a typical peak intensity, is given as the solid green line, with a scale indicated on the right vertical axis.}
\label{fig:franck-condon}
\end{figure}

The $R$ dependence of the square root of the ionization rate $W_{\mathrm{PPT}}$, as it enters Eq.\,(\ref{state_selected_single_ion_rate}), is displayed in Fig.\,\ref{fig:franck-condon}, panels (c) and (d) together with a pair of H$_2^+$ vibrational wavefunctions $\chi_{v_+}(R)$, with $v_+=1$ in (c) and $v_+=2$ in (d). The FC distribution, $f^{FC}(v_+)$ is given by the histogram of panel (a), while that in panel (b) gives an example of non-Franck-Condon (non-FC) distribution, $f^{nFC}(v_+)$, as given in Eq.\,(\ref{Pv_plus_nonFC}) with $\Gamma_0^{v_+}$ therein defined by  Eq.\,(\ref{state_selected_single_ion_rate}) using $W_{\mathrm{PPT}}$. The distribution shown in panel (b) is  for the specific case of a  $\lambda=800$\,nm, $I=1.6\times10^{14}$\,W/cm$^2$ linearly polarized pulse.

The first general observation that can be made when comparing the FC with the non-FC distributions is that the maximum population is shifted from the $v_+=2$ level to the $v_+=1$ level, the ground $v_+=0$ population increasing also with respect to its FC value. The second observation is that the distribution stops at a lower value of  $v_+$ than in the FC distribution: While the FC distribution in panel (a) extends to the last  $v_+$ level ($v_+=18$) supported by  the ground-state potential energy curve of H$_2^+$, the non-FC  distribution in panel (b) acquires negligible values past $v_+ \simeq 10$. The quantitative details of these deviations with respect to the vertical FC ionization results depend on the specific values of the laser pulse parameters.  We will give later a more detailed discussion of how these variations with respect to the FC distribution depend on the field polarization. For now, it suffices to say that the non-FC distribution that the MO-PPT ionization model affords is more compact, more restricted to the low-lying vibrational levels of the cation.

We can understand these salient traits of the non-Franck-Condon distributions by referring to panels (c) and (d) of Fig.\,$\ref{fig:franck-condon}$. One first  notes that a rather fast variation from almost zero to 0.7 is observed for $W_{\mathrm{PPT}}^{1/2}$ as $R$ goes from 0 to ca. $3$\,a.u. This has a double effect: First, recalling Eq.\,(\ref{state_selected_single_ion_rate}), we see that  with respect to high $v_+$ vibrational wavefunctions of H$_2^+$, it  filters out almost completely the left turning point amplitudes of this wavefunction, which normally (i.e. in the FC case) would presents some overlap with the $\chi_{0}(R)$ H$_2$ ground vibrational state, while the amplitudes of this  high $v_+$ wavefunction found at the right turning point lies completely  outside the range of $\chi_{0}(R)$ and  gives a zero overlap with it. The consequence is that, from a broad vibrational population distribution extending up to $v_+=18$ as obtained for the vertical FC ionization, we now have, with the non trivial dependence of $W_{\mathrm{PPT}}$ on $R$,  a narrower distribution  (up to $v_+ \simeq 10$).

The fact that this non-FC distribution is peaked at $v_+=1$ rather than at $v_+=2$ can also be understood by this rise of $W_{\mathrm{PPT}}^{1/2}$ in the region where $\chi_{0} (R)$ is peaking, though things here depend on a rather delicate balance between various portions of the variation of $W_{\mathrm{PPT}}^{1/2}$ around its value at the equilibrium internuclear distance $R_e$ of the neutral molecule. In the interval of $R$ where $\chi_{0} (R)$ overlaps well with  $\chi_{v_+=1} (R)$ (traced in orange in panel (c)), \emph{i.e.} on the right of the equilibrium geometry, $W_{\mathrm{PPT}}^{1/2}$ is larger than its value at $R_e$. Thus the overlap between the $v_+=1$ state and the H$_2$ ground vibrational wavefunction is enhanced with respect to the FC situation. The same thing can be said for the overlap between the cation ground vibrational wavefunction. This nodeless $\chi_{v_+=0} (R)$ wavefunction is centered at the node of the $\chi_{v_+=1} (R)$ function just discussed and its overlap with the H$_2$ ground vibrational state $\chi_{0} (R)$ lies in the region where $W_{\mathrm{PPT}}^{1/2}$ is larger than its value at $R_e$. As a consequence, the weights of the $v_+=0$ and $v_+=1$ vibrational states are increased when using the more realistic PPT approach as compared to the FC approximation. 

The case of the vibrational state $v_+=2$ is more delicate: In the interval of $R$ where $\chi_0(R)$ overlaps well with $\chi_{v_+=2}(R)$ (traced in orange in panel (d)), $W_{\mathrm{PPT}}^{1/2}$ is smaller than its value at $R_e$  on the left of this equilibrium geometry, diminishing the contribution to the overlap  there. The subsequent rise of $W_{\mathrm{PPT}}^{1/2}$ in the region of the first maximum of $\chi_{v_+=2}(R)$ contributes to enhance the (positive) overlap there. However, the continued rise of $W_{\mathrm{PPT}}^{1/2}$ past this range enhances the previously small negative overlap between the tail of $\chi_{0}(R)$ and the negative lobe of $\chi_{v_+=2} (R)$. All this gives a slight decrease in the overlap between the two functions when the $R$-dependent $W_{\mathrm{PPT}}^{1/2}$ is inserted.

\begin{figure}[t!]
\centering
\includegraphics[width=0.99\linewidth]{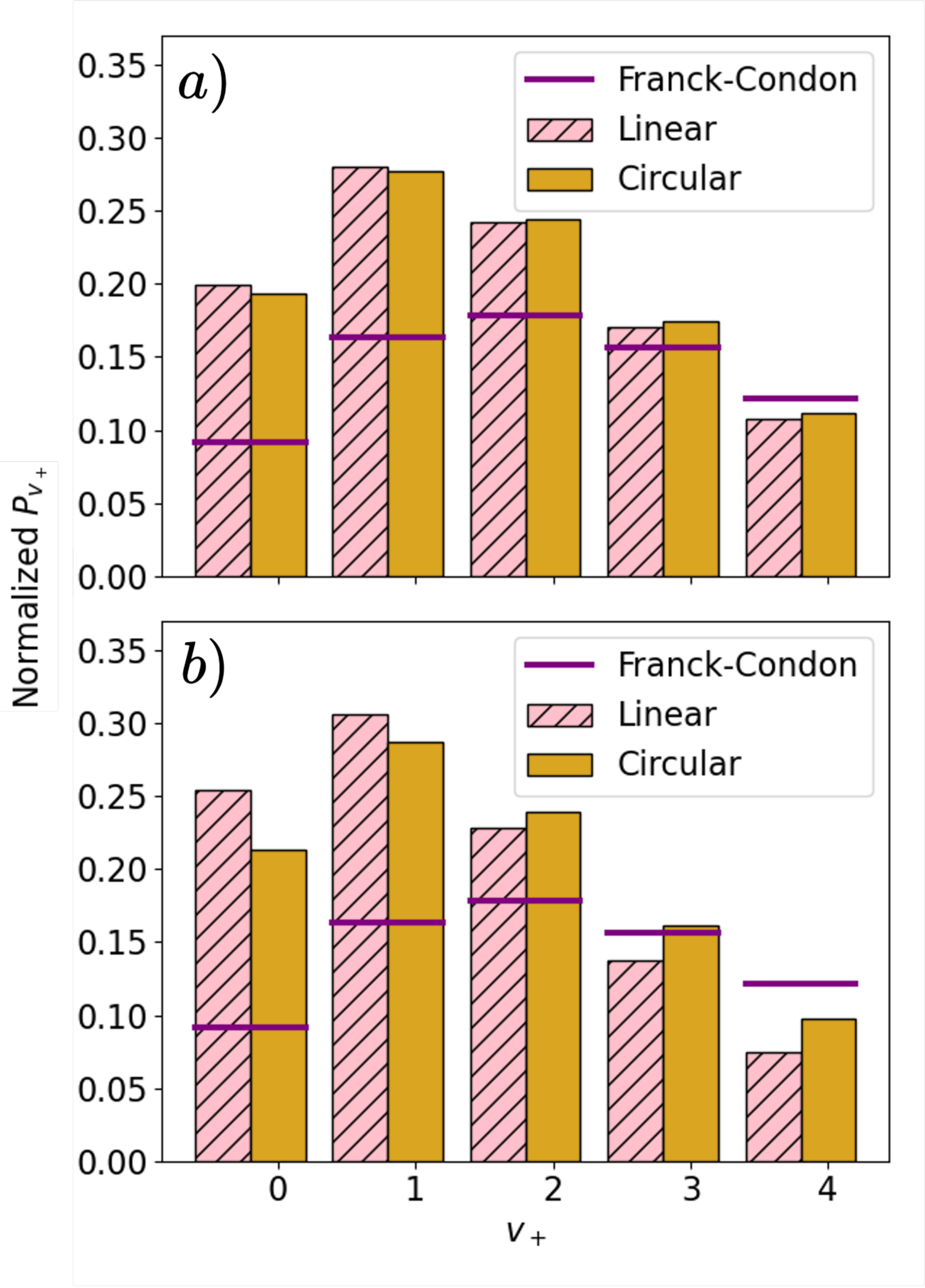}
\caption{Non-FC final H$_2^+$ populations using linearly (striped pink bars) and circularly (full brown bars) polarized fields, calculated at a peak intensity of $I=1.6\times10^{14}$\,W/cm$^2$ by the MO-PPT method. The upper and lower panels are for comparing wavelengths of $(a)$ 266 nm over 36 optical cycles and $(b)$ 800 nm over 12 optical cycles. FC projections are indicated by horizontal purple lines.}
\label{nonFC}
\end{figure}

A closer look at how the non-FC character affects the lowest $v_+$ states' populations is given in Fig.\,\ref{nonFC}, for two values of the wavelengths, $266$\,nm and $800$\,nm. The field intensity is $1.6\times10^{14}$\,W/cm$^2$. The histogram with striped pink bars is for the linear polarization case, while that with full brown bars is for the case of a circularly polarized pulse. The horizontal purple line drawn at each value of $v_+$ marks the expected height of the population for that level in the FC distribution. We see that there is depletion of population with respect to the FC distribution  starting from $v_+=4$. The $v_+=2$ population actually increases, but less so than the $v_+=0$ and $1$  populations. There is little difference between linearly and circularly polarized fields in the MPI regime ($\lambda=266$\,nm, panel (a)). The non-FC character seems a little stronger for linearly polarized fields in the TI regime ($\lambda=800$\,nm, panel (b)). 

In summary, the ionization of H$_2$ within a MO-PPT model description, enhances the populations of  the lowest levels $v_+=0-4$ of H$_2^+$ to the detriment of levels close to the dissociation limit. A number of remarks ought to be made at this point. First, although the above results were obtained for rather long pulses, they are representative of the vibrational excitation of the cation as it is formed, in an attosecond time-scale, as the MO-PPT semi-classical model sees the time as merely a parameter governing the instantaneous value of the field under which the molecular ionization is produced. Basically, the consideration of the long pulse is irrelevant in the discussion of the non-FC character of the ionization, as it merely amplifies the collected ionization signal while the non-FC character is defined relative to the total final ionization probability, (see Eqs.\,(\ref{Pv_plus_nonFC}, \ref{Pv_plus_FC})). Thus, the $v_+$-state population distributions shown here represent the vibrational wave packet of the cation as it is formed. {This absence of dependence of the calculated non-FC distributions with the pulse duration was confirmed by calculations performed with shorter (8\,fs FWHM) and longer (32\,fs FWHM) pulses at the 3 considered wavelengths (266, 400 and 800\,nm). This distribution is thus} to be viewed as the initial state of a subsequent vibrational dynamics, be it with or without an external field.

Imagine for example that the external field is the probe pulse in a pump-probe experiment. The outcome of this vibrational dynamics, often discussed by assuming a FC preparation of the molecular ion, will be affected by this non-FC character. In the IR spectral region, a dynamical control of the dissociation probability can be exerted by tuning the carrier-envelope phase of the probe pulse, by a mechanism called Dynamical Dissociation Quenching (DDQ) \cite{JCP_108_3974, Abou-Rachid_DDQ_2001}. The non-FC wave packet would modify the optimized parameters for this type of control. In the UV-Visible region, an intense probe pulse gives rise to ATD signals in the proton kinetic energy release (KER) spectrum, with possible signatures of Bond-Softening and/or Vibrational Trapping effects \cite{giusti-suzor_mies_vibrational_trapping_1992}. All this depends delicately on the content of the initial vibrational wave packet. A non-FC preparation of this wave packet will surely be manifest in these spectra. 

The last comment to be made on these non-FC ionization considerations is that the semi-classical expressions of the ionization rate depend on $R$ only through the ionization potential, as they are basically adapted from a theory of atomic ionization. By considering the core potential as felt by the ionizing electron as a single-centered one, MO-ADK and MO-PPT, as well as  CS-ADK, the  recent adaptation of ADK theory to complex systems \cite{Hu_2013}, ignore the LCAO (Linear Combination of Atomic Orbitals) form of the molecular orbital out of which the ionization takes place. A study that takes into account the form of this potential\,\cite{R_Murray_U_Waterloo}, using a reformulation of the tunneling ionization theory\,\cite{Murray_Liu_Ivanov_PhysRevA.81.023413} do reveal this dependence that would further enhance the non-Franck-Condon character of the cation initial vibrational wave packet. Various considerations, of short-time projections of the initial (molecular) orbitals onto the ionization continuum, or of exact molecular symmetry conservation in the case of ionization under a perpendicularly polarized field\,\cite{Mol_Phys_LIED_symmetry_doi:10.1080/00268976.2017.1317858}, indicate the existence of diffraction-like factors in the exact ionization probability amplitude, the $R$-dependence of which would contribute to a stronger non-FC behaviour, in particular for a high ionized electron momentum. This aspect of the non-Franck-Condon strong field ionization is yet to be explored using a more detailed, albeit numerically exact description of the strong-field ionization.

\section{Conclusion}\label{concl}

In this paper, we consider the dynamical process of laser-induced strong field ionization of H$_2$, treated within semi-classical models which include the internuclear degree of freedom as a mere parameter. We verified that for laser peak intensities not exceeding a few $10^{14}$\,W/cm$^2$ the single (first) ionization is by far the dominant process in the tunnel ionization regime. For those intensities the first ionization process can be considered by itself, separated from the slower vibrational motions, thus allowing us to concentrate on the ionization dynamics. Concerning the models, we refer to semi-classical treatments comparing the relative performances of MO-ADK and MO-PPT approximations. In that respect, we show that the popular MO-ADK method performs well in the tunnel ionization (TI) regime, that is for the corresponding Keldysh parameter $\gamma \ll 1$, \emph{i.e.} for high intensities and long wavelengths. The MO-PPT formula is characterized by a more extended validity domain, at least partially covering the multiphoton ionization (MPI) regime with $\gamma \gg 1$. As typical examples of the TI and MPI regimes, we considered two wavelengths, one in the near IR domain (800\,nm) and the other one in the UV spectral region (266\,nm). We compared the results for the two types of field polarizations: linear and circular. For any wavelength, the threshold intensity for the H$_2^+$ population saturation is lower for the case of a circular polarization as compared to the linear polarization case. This is mainly a fluence effect. We also noted that the slope of the rise of the ionization probability vs. intensity curve is higher for $\lambda=800$\,nm than for $\lambda=266$\,nm. This is coherent with the identification of the first case as typical of the TI regime, and the second of the MPI regime.

Concerning the vibrational populations of the ion as it is formed, we have shown that the strong field ionization of H$_2$ does not follow the Franck-Condon principle, contrary to the simplified view of an ultrafast ionization mechanism with the nuclear motions frozen on the ionization time scale. Indeed, in this simplified view the vibrational initial state of the neutral molecule would simply be transported vertically onto the cation's ground-state manifold. The deviation from a Franck-Condon behaviour is to be understood in terms of the $R$-dependence of the ionization rate, which increases rapidly with the internuclear distance, whatever the polarization. The resulting so-called non-FC distributions are peaked at lower vibrational levels than the FC ones, and are more limited in $v_+$ range than the Franck-Condon one, levels above $v_+=4$ being only weakly populated.

It is worthwhile noting that non-FC distributions have already been experimentally observed in the literature using a 800\,nm excitation in linear polarization, and interpreted with a theoretical approach at the ADK level of approximation\,\cite{PhysRevLett.92.163004}. This study concluded also that the ionization process was favoring the production of the lowest $v_+$ levels, but the comparison cannot be pushed further, not only because our study is performed at the PPT level of approximation, but also because the individual vibrational levels of H$_2^+$ referred to in Ref.\,\cite{PhysRevLett.92.163004} take into account the laser-induced modifications of the two field-dressed potential energy curves of the $1s\sigma_g$ and $2p\sigma_u$ states of the cation. As such, the distribution is necessarily non-FC, not because of a non-vertical ionization, but because the distribution is one over a different basis of vibrational states.

The study we have presented here, describing the creation of a vibrational wave packet in the electronic ground state of the molecular ion H$_2^+$, is a first step towards the development of a more complete model that will take into account the subsequent nuclear dynamics, by solving the time-dependent Schrödinger equation describing the dynamics of the nuclear wave packet thus formed in the electronic state $1s\sigma_g$ and coupled to the first excited and dissociative electronic state $2p\sigma_u$ of the H$_2^+$ ion. This will complete the inclusion of a semi-classical MO-ADK or MO-PPT scheme in the full description of the dissociative ionization process leading to a useful observable such as the kinetic energy distribution of protons emitted in the dissociative continuum associated with the asymptotic [H(1s)\,+\,H$^+$] channel. This approach should allow an accurate and realistic evaluation of the low energy part of these proton spectra, while the description of higher kinetic energies will require the consideration of the double ionization process leading to the [H$^+$\,+\,H$^+$] Coulomb Explosion occurring for stronger laser fields. Work is thus in progress in our group to treat the complete dynamics that can take place, from the ionization of H$_2$ to its Coulomb Explosion, based on a semi-classical approach for the ionization and a fully quantum description for the vibrational dynamics.

\backmatter

\bmhead{Supplementary information}
{Results of calculations with varying pulse durations, as mentioned and discussed in section \ref{3.1}, are shown in a separate Supplementary Information file.}

\bmhead{Acknowledgments}
This work has been performed within the French GDR UP number 3754 of CNRS. Jean-Nicolas Vigneau is grateful to the French MESRI (French Ministry of Higher Education, Research and Innovation) for funding his PhD grant through a scholarship from EDOM (Ecole Doctorale Ondes et Matière, Université Paris-Saclay, France). JNV also acknowledges partial funding from the Choquette Family Foundation - Mobility Scholarship and the Paul-Antoine-Giguère Scholarship.

\section*{Declarations}

\subsection*{Competing interests}
All authors certify that they have no affiliations with or involvement in any organization or entity with any financial interest or non-financial interest in the subject matter or materials discussed in this manuscript.

\subsection*{Data availability statement}
The datasets generated during and/or analysed during the current study are available from the corresponding author on reasonable request.

\subsection*{Code availability}
The Fortran codes developed by us and used in this work are available on request from the authors.

\subsection*{Authors' contributions}
All authors contributed to the study conception and design. The numerical simulations were performed by Jean-Nicolas Vigneau. The first draft of the manuscript was written by Osman Atabek and all authors commented on previous versions of the manuscript. All authors read and approved the final manuscript.

\pagebreak

~

\pagebreak

\onecolumn

\section*{Supplementary Information}

\subsection*{Effect of the pulse duration on the ionization probability}

To illustrate the effect of pulse duration, we present here, for $\lambda=800$\,nm, the results of calculations made for a pulse of  8\,fs FWHM, corresponding to half the pulse duration considered in the main text, and of 32\,fs FWHM, i.e. the double of the pulse duration in the main text. The results for the shorter pulse, 8\,fs FWHM (6 optical cycles), and those for the longer pulse, 32\,fs FWHM (24 optical cycles) are given in Fig.\,\ref{fig:pop_1stion-lin} for a linearly polarized field, and in Fig.\,\ref{fig:pop_1stion-circ} for a circularly polarized field.

\makeatletter 
\renewcommand{\thefigure}{S\@arabic\c@figure}
\makeatother

\begin{figure*}[ht!]%
    \centering
    \includegraphics[width=0.99\linewidth]{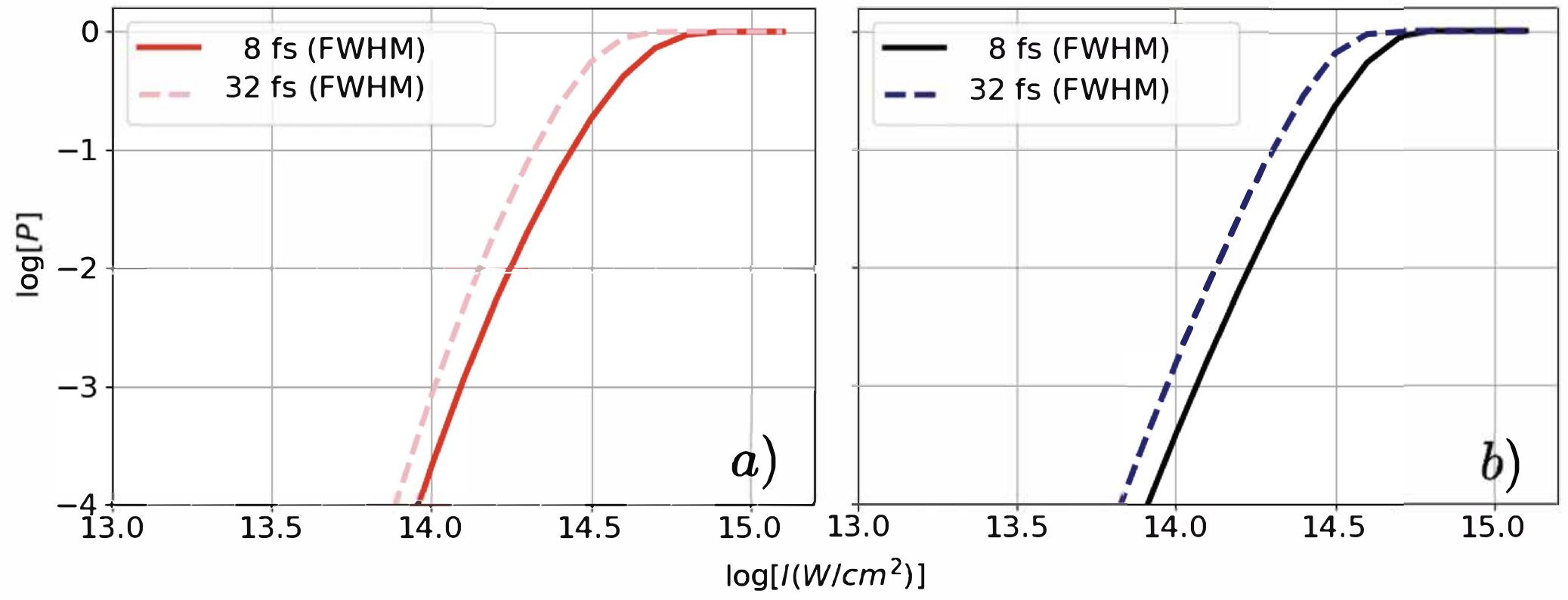}
    \caption{Final ionization probability of H$_2$ (in logarithmic scale) using a linearly polarized field of peak intensity $I$ (in logarithmic scale), as given by (a) MO-ADK (red and pink) and (b) MO-PPT (black and blue) methods, with wavelengths of 800 nm over 6 optical cycles (8\,fs FWHM; full red and black lines) and 24 optical cycles (32\,fs FWHM; dashed pink and blue lines).}
    \label{fig:pop_1stion-lin}
\end{figure*}

\begin{figure}[ht!]%
    \centering
    \includegraphics[width=0.99\linewidth]{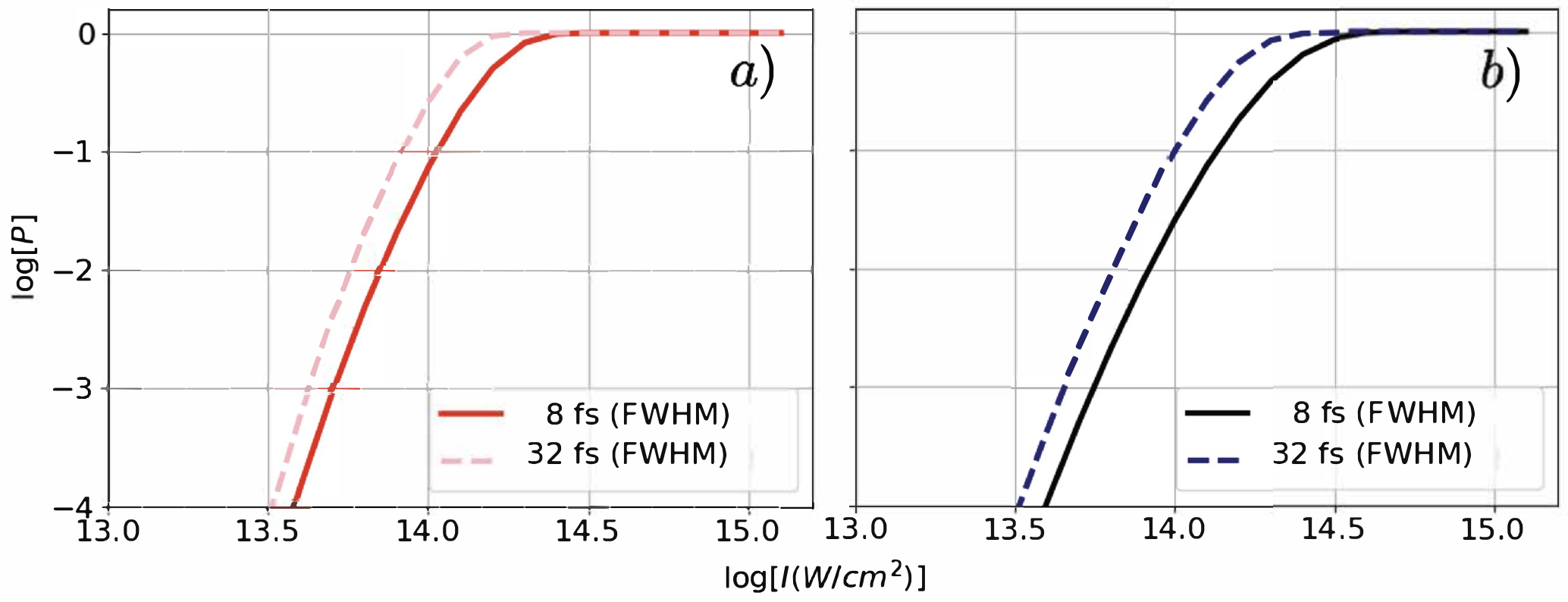}
    \caption{Final ionization probability of H$_2$ (in logarithmic scale) using a circularly polarized field of peak intensity $I$ (in logarithmic scale), as given by (a) MO-ADK (red and pink) and (b) MO-PPT (black and blue) methods, with wavelengths of 800 nm over 6 optical cycles (8\,fs FWHM; full red and black lines) and 24 optical cycles (32\,fs FWHM; dashed pink and blue lines).}
    \label{fig:pop_1stion-circ}
\end{figure}

\noindent Compared with panel (c) of Fig.\,4 of the main text, it is clearly seen  that  the ionization probability at $t_f$ decreases (increases) globally with the pulse compression (stretching), and the saturation intensity (where $P_{ion}(t_f)=1$) is shifted to higher (lower) values.
\medskip

\begin{figure}[ht!]%
    \centering
    \includegraphics[width=0.99\linewidth]{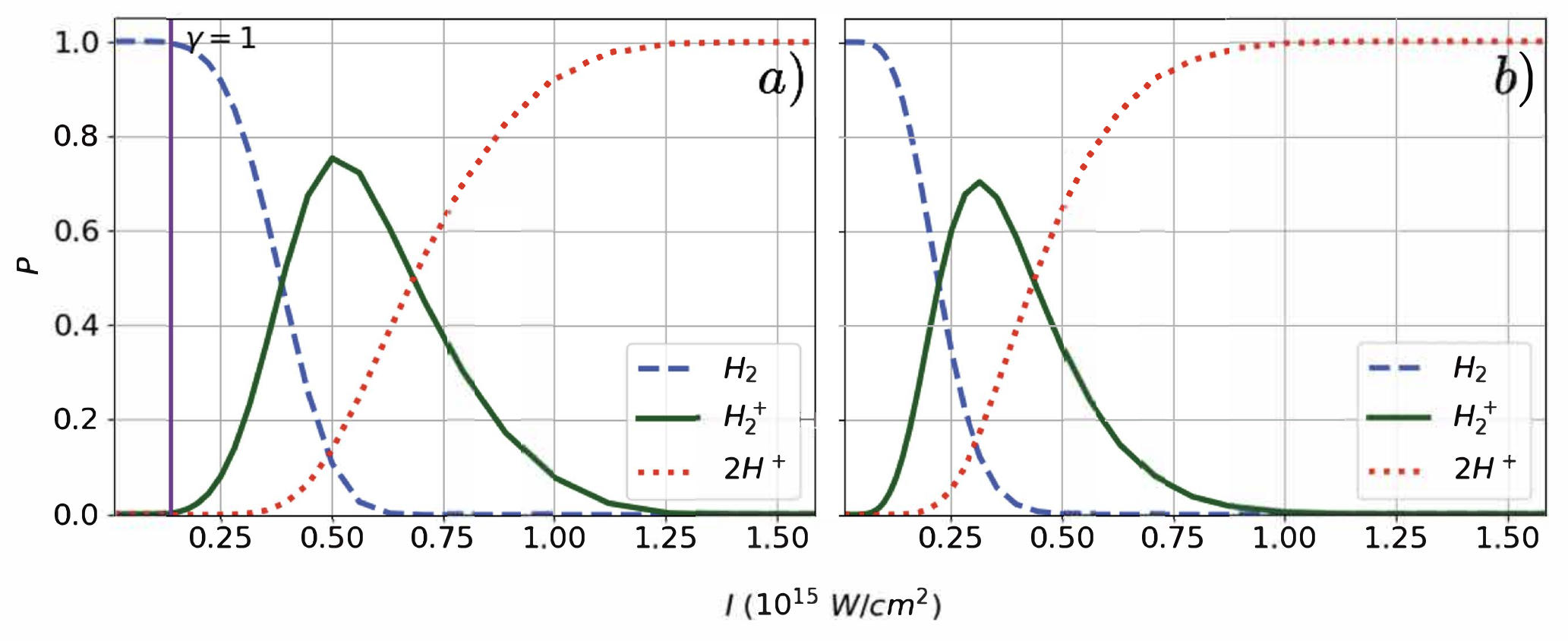}
    \caption{Final populations of H$_2$ ($P_{\mathrm{H}_{2\!}}(t_f)$; dashed blue line), H$_2^+$ ($P_{\mathrm{H}_2^{+\!}}(t_f)$; full green line) and H$^+$\,+\,\,H$^+$ ($P_{\mathrm{CE}}(t_f)$; Coulomb Explosion, red dotted line), using a 800\,nm (a) linearly and (b) circularly polarized field, and the MO-PPT approach for a sin$^2$ pulse envelope of total duration 16\,fs corresponding to 6 optical cycles (8\,fs FWHM). The vertical purple line in panel (a) indicates the value of the {intensity at which the} Keldysh parameter is $\gamma=1$.}
    \label{fig:pop_tot}
\end{figure}

\noindent The effect of the pulse duration shortening or stretching on the second ionization follows the same trend, causing a shift of the H$_2^+$ (or  H$_2^{++}$) apparition curve(s) to a higher (lower) intensity, i.e. to the right (left) of its value in Fig. 3 of the main text, in both the linear and circular polarization results. Fig.\,\ref{fig:pop_tot} shows the results for the pulse compression case (6 optical cycles, 8\,fs FWHM), using the same layout as Fig.\,3 of the main text.

\end{document}